\documentclass[conference]{IEEEtran}
\IEEEoverridecommandlockouts

\usepackage{cite}
\usepackage{amsmath,amssymb,amsfonts}
\usepackage{algorithm}
\usepackage{algorithmic}
\usepackage{graphicx}
\usepackage{textcomp}
\usepackage{xcolor}
\usepackage{colortbl}
\usepackage{enumitem}
\usepackage{xspace}
\usepackage{pifont}
\usepackage{url}
\usepackage{microtype}

\makeatletter
\def\section{\@startsection{section}{1}{\z@}{0.55ex plus 0.8ex minus 0.3ex}%
{0.4ex plus .4ex minus 0ex}{\normalfont\normalsize\centering\scshape}}%
\def\subsection{\@startsection{subsection}{2}{\z@}{0.55ex plus 0.8ex minus 0.3ex}%
{0.35ex plus .3ex minus 0ex}{\normalfont\normalsize\itshape}}%
\makeatother

\def\IEEEtitletopspace{0pt}
\makeatletter
\def\@IEEEauthorblockconfadjspace{-0.7em}
\def\@IEEENORMtitlevspace{0.75\baselineskip}
\def\@IEEEMINtitlevspace{0.5\baselineskip}
\makeatother

\def\BibTeX{{\rm B\kern-.05em{\sc i\kern-.025em b}\kern-.08em
    T\kern-.1667em\lower.7ex\hbox{E}\kern-.125emX}}

\newcommand{\name}{\textsc{BIDENT}\xspace}

\renewcommand{\baselinestretch}{0.84}

\graphicspath{{figures/}}

\begin{document}

\title{\name: Heterogeneous Operator-Level Mapping for Efficient Edge Inference}

\author{\IEEEauthorblockN{Hoseok Kim$^{\dagger}$, Arghadip Das$^{\dagger}$, Soumendu Ghosh$^{\ddagger}$, Arnab Raha$^{\ddagger}$, Vijay Raghunathan$^{\dagger}$}
\IEEEauthorblockA{$^{\dagger}$Purdue University \quad $^{\ddagger}$Intel Corporation \\
$^{\dagger}$\{kim5396, das169, vr\}@purdue.edu, \quad $^{\ddagger}$\{soumendu.ghosh, arnab.raha\}@intel.com}}

\maketitle
\bstctlcite{BSTcontrol}

\begin{abstract}
Modern edge System-on-Chips (SoCs) integrate increasingly heterogeneous processing units (PUs) such as CPUs, GPUs, and NPUs, yet current inference stacks still map entire models to a single PU, leaving significant performance and energy efficiency on the table. This limitation is exacerbated by emerging neural architectures such as state-space models (SSMs), Kolmogorov--Arnold networks (KANs), and multi-stage vision-language-action (VLA) pipelines, whose diverse operator characteristics are not uniformly suited to any single PU.

We present \name, a unified operator-level orchestration framework for heterogeneous edge inference that maps individual operators to the most suitable PU based on profiled execution characteristics. \name formulates operator-to-PU assignment as a shortest-path problem over a weighted execution graph, enabling efficient and principled scheduling under the profiled cost model for both latency and energy minimization objectives. Unlike prior work that relies on model-specific heuristics or coarse-grained partitioning, \name provides a model-agnostic framework that jointly supports sequential execution, intra-model parallelism across independent operators, and concurrent co-scheduling of two inference requests within a single formulation. Evaluated on an Intel Core Ultra SoC across ten model families and 190 concurrent model pairs, \name achieves up to 1.60$\times$ single-model speedup, reduces energy by 9.2\% on average, and delivers a 2.28$\times$ geometric-mean concurrent speedup over a naive multi-PU placement baseline. \looseness=-1


\end{abstract}


\section{Introduction}
\label{sec:introduction}

On-device AI has evolved from narrow perception tasks to workloads that demand sustained reasoning, multimodal fusion, and real-time decision-making, and modern edge System-on-Chips (SoCs) accordingly integrate heterogeneous processing units (PUs) such as CPUs, GPUs, and NPUs (Neural Processing Units). \looseness=-1

Concurrently, edge neural network architectures are diversifying beyond the Transformers and CNNs (Convolutional Neural Networks) that deployment stacks are designed around.
Emerging models introduce operators with distinct compute requirements, such as FFT-based long convolutions (Hyena~\cite{hyena}), control-heavy spline evaluations (KANs~\cite{kan}), and selective scan kernels (Mamba~\cite{mamba}), while multi-stage vision-language-action (VLA) pipelines ({$\pi_{0.5}$}~\cite{phi05}) chain encoders, decoders, and denoising stages with distinct operator mixes.
Such operators can suffer order-of-magnitude latency and energy penalties on a mismatched PU~\cite{hkn}, and edge systems increasingly serve multiple inference streams concurrently, amplifying the cost of suboptimal PU assignment. \looseness=-1

Despite this diversity, current deployment frameworks map an entire model's execution graph to a single target PU.
This monolithic approach has three limitations.
It ignores per-PU capability differences, forcing every operator onto one PU regardless of affinity~\cite{hkn,das2025grannite,das2025xamba,raha2025llm}.
It leaves hardware underutilized, serializing data-independent operators that could execute concurrently across PUs.
It also exposes unsupported operators, which fall back to inefficient paths, silently substitute approximations, or fail to compile altogether~\cite{hkn}. \looseness=-1



\begin{figure}[t]
    \centering
    \includegraphics[width=0.92\columnwidth]{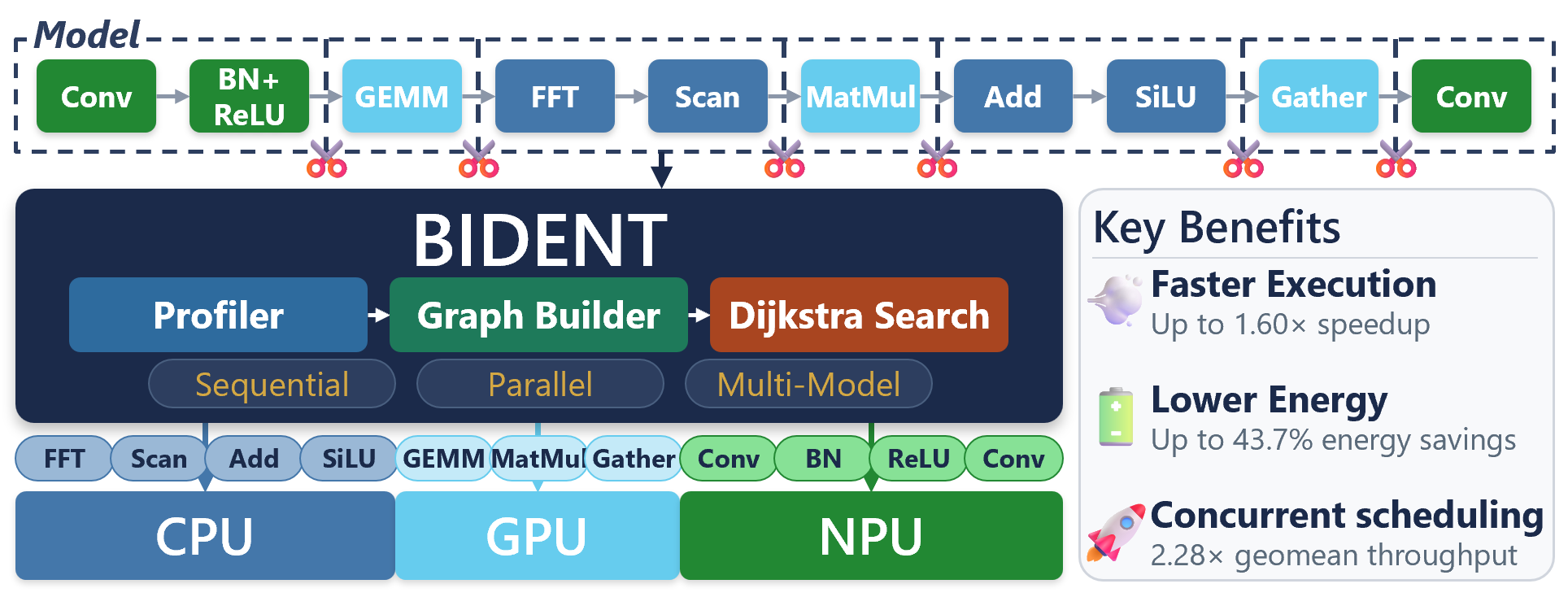}
    \vspace{-7pt}
    \caption{\name overview.}
    \label{fig:overview}
    \vspace{-8pt}
\end{figure}

To address these limitations, we propose \name, an operator-level orchestration framework for heterogeneous edge SoCs (Fig.~\ref{fig:overview}). \name profiles each fused operator of the input models on every PU, encodes the measured costs as a weighted execution graph, and derives a static schedule by shortest-path search under a latency or energy objective. One formulation covers sequential, intra-model parallel, and two-request concurrent execution.
\looseness=-1

The key contributions of this work are: \looseness=-1
\begin{itemize}[leftmargin=*,nosep]
    \item {\bf A principled graph-search formulation with scoped optimality.} We cast operator-to-PU assignment as a shortest-path problem over a weighted execution graph, reducing combinatorial scheduling to a polynomial-time optimization that is exact for sequential chains under the profiled cost model, for both latency and energy objectives.\looseness=-1
    \item {\bf A common orchestration framework across execution regimes.} \name supports sequential heterogeneous mapping, intra-model parallel execution of independent operators, and two-request concurrent co-scheduling within a single framework. \looseness=-1
    \item {\bf Comprehensive evaluation across diverse models and workloads.} Across ten model families and 190 concurrent model pairs, \name achieves up to 1.60$\times$ single-model speedup with intra-model parallelism and a 3.42$\times$ geometric-mean speedup in concurrent settings, 2.28$\times$ over naive multi-PU placement, demonstrating the importance of operator-level orchestration. \looseness=-1
\end{itemize}
\section{Background and Motivation}
\label{sec:background_motivation}


\subsection{Diverse Neural Network Architectures}
\label{subsec:diverse_architectures}

Modern edge SoCs combine CPUs, GPUs, and NPUs that share memory but favor different workloads~\cite{apple_m5_2025,intelcoreultra}. CPUs suit irregular and sequential computation, GPUs highly parallel GEMMs, and NPUs MAC-dense convolutions, so CNN convolutions map well to NPUs, the best PU for Transformer GEMMs varies with phase and hardware~\cite{agent.xpu}, and Hyena FFTs~\cite{hyena}, Mamba recurrences~\cite{mamba}, and KAN splines~\cite{kan} often favor CPUs. No single PU is consistently optimal. \looseness=-1

Practical models combine operators with different PU affinities. Transformers mix GEMM-heavy attention with non-linear layers, hybrid models add SSM or Hyena blocks, and VLA pipelines chain partially independent stages. Concurrent workloads further create contention and idle capacity under static single-PU mapping. \looseness=-1


\begin{figure}[t]
    \centering
    \includegraphics[width=\columnwidth]{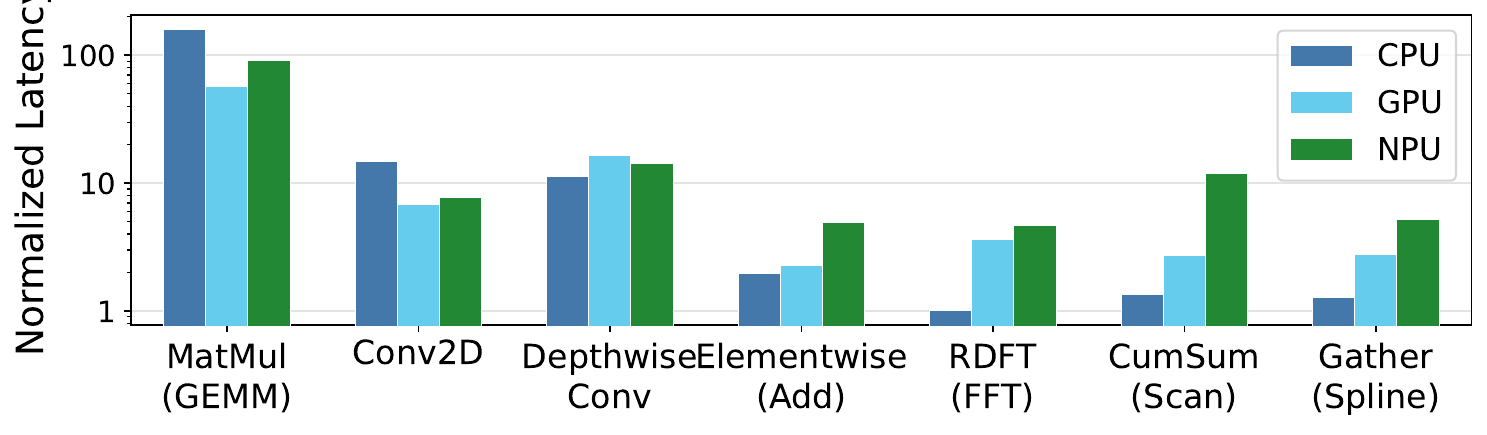}
    \vspace{-18pt}
    \caption{Execution latency of seven representative operators across CPU, GPU, and NPU on an Intel Core Ultra SoC, measured with 200 inferences at FP16. All bars share one normalization base, the single fastest measurement (\texttt{RDFT} on the CPU), so heights are comparable across operators and PUs.}
    \label{fig:operator_type_comparison}
    \vspace{-6pt}
\end{figure}

\subsection{PU Capability Variation}
\label{subsec:capability_variation}
\noindent\textbf{Observation 1: Best PU depends on operator type.}
Fig.~\ref{fig:operator_type_comparison} profiles seven representative operators. The GPU is fastest for \texttt{MatMul} and \texttt{Conv2D} (2.8$\times$ and 2.2$\times$ over the CPU), whereas the CPU leads everywhere else. The NPU suffers most on the non-GEMM \texttt{RDFT}, \texttt{CumSum}, and \texttt{Gather} (4.7--8.8$\times$ over the CPU), which fall outside its MAC-centric datapath, and dispatch overhead even costs it supported operators such as \texttt{DWConv} at the tested shapes. \looseness=-1

\begin{figure}[t]
    \centering
    \includegraphics[width=\columnwidth]{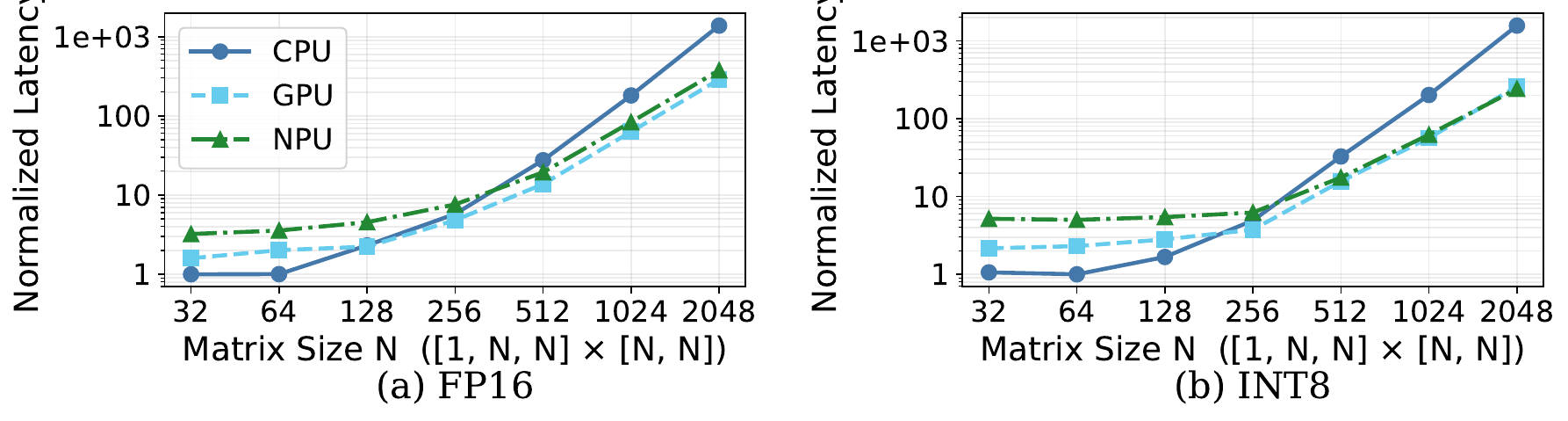}
    \vspace{-18pt}
    \caption{Execution latency of square \texttt{MatMul} of varying matrix size across CPU, GPU, and NPU on an Intel Core Ultra SoC, at FP16~(a) and INT8~(b) (each normalized to its fastest measurement). \looseness=-1}
    \label{fig:operand_size_comparison}
    \vspace{-6pt}
\end{figure}

\noindent\textbf{Observation 2: Best PU depends on operand size.}
Fig.~\ref{fig:operand_size_comparison} sweeps square \texttt{MatMul} for $N \in \{32,\ldots,2048\}$. The CPU leads for small matrices, the GPU takes over at $N=128$ for FP16 and $N=256$ for INT8 (up to 4.8$\times$ over the CPU), and the NPU wins only at $N=2048$ in INT8, where its compute density finally offsets its dispatch overhead. \looseness=-1



\subsection{SoC Underutilization}
\label{subsec:pu_underutilization}

Single-PU serialization also leaves the remaining PUs idle while data-independent subgraphs execute sequentially.\looseness=-1


\noindent\textbf{Observation 3: Parallel execution recovers idle PU capacity.}
We run three independent operator pairs (\texttt{MatMul}$\|$\texttt{Conv2D}, \texttt{MatMul}$\|$\texttt{CumSum}, \texttt{Conv2D}$\|$\texttt{DWConv}) across all six ordered PU assignments. GPU$\|$CPU achieves the highest speedup for every pair ($1.41\times$, $1.38\times$, and $1.46\times$ over the best serial single PU), whereas an unsuitable assignment underperforms the serial baseline, so parallelism must be combined with operator-aware PU selection, a principle that extends to co-scheduling concurrent models (Section~\ref{subsec:parallel_results}). \looseness=-1

Together, Observations~1--3 motivate a system-level orchestration framework that (1) maps each operator to the PU best suited for its type and operand size, (2) exploits inter-operator parallelism for data-independent subgraphs, and (3) co-schedules concurrent inference streams across idle PUs, the three core capabilities of \name. \looseness=-1

\section{\name Framework}
\label{sec:framework}

\name is an operator-level orchestration framework for heterogeneous edge SoCs. Given one or two neural network model instances and the target SoC's processing units (PUs), it produces a static schedule specifying operator-to-PU assignments, execution order, and synchronization. \name minimizes either end-to-end latency or total package energy in every supported mode. Here, an \textit{operator} denotes a compiler-fused group of primitive operations (e.g., \texttt{Conv--BN--ReLU}), following~\cite{axonn,haxconn}. \name comprises a per-operator profiler, a weighted execution-graph builder, and a shortest-path search engine, applied in three offline stages of profiling, graph construction, and search.\looseness=-1

Our implementation uses OpenVINO~\cite{openvino} for per-PU compilation and inference. The formulation is backend-agnostic and adapts to TensorRT~\cite{tensorrt}, ONNX Runtime~\cite{onnxruntime}, or QNN~\cite{qnn}, provided the backend exposes fused-operator boundaries and per-PU sub-model compilation.\looseness=-1

\subsection{Profiler}
\label{subsec:profiling}

The profiler records, for each supported operator--PU assignment, host-to-device (H2D) transfer latency, kernel-dispatch latency, hardware-counter execution time, and device-to-host (D2H) transfer latency, together with each model's average package power during sustained inference on each PU. We denote the dispatch-plus-execution latency of operator $O_i$ on PU $P_j$ by $w(O_i,P_j)$ and the model's sustained power on $P_j$ by $\bar{p}_j$, while transfer costs are stored separately.

Profiling is performed offline once per model--SoC pair. The profiler additionally measures pairwise cross-PU slowdown factors for intra-model parallelism and, for two concurrent requests, barrier-synchronized same-PU and cross-PU operator-pair co-execution costs. In our OpenVINO implementation, \name derives a single fused-operator partition from per-operator \texttt{PERF\_COUNT} profiling on the NPU, falling back to a supported PU for the rare model that cannot compile there, and reuses the partition boundaries across all PUs. Each operator is extracted with its original shape and layout, compiled independently for every supported PU, and measured, capturing shape-dependent PU affinity. Unsupported operator--PU assignments are omitted, ensuring that the execution graph contains only feasible mappings.\looseness=-1

\subsection{Execution Graphs and Mapping Search}
\label{subsec:execution_graph}

The execution graph builder converts the profiled costs and model dependencies into weighted DAGs whose paths encode feasible schedules. \name then applies standard shortest-path algorithms to these graphs; its contribution lies in the operator-level formulation and empirically measured cost model.\looseness=-1

\noindent\textbf{Sequential execution graph.}\label{subsubsec:sequential_graph}
For a linear chain $O_1\rightarrow\cdots\rightarrow O_N$, \name creates a node $v_{i,j}$ for each supported assignment of operator $O_i$ to PU $P_j$, together with virtual start and end nodes $s$ and $t$. The node cost is the profiled cost under the selected objective, either latency $w(O_i,P_j)$ or measured energy.\looseness=-1

Edges from $s$ to $v_{1,j}$ contain the initial H2D cost required by $P_j$. Between consecutive operators, execution on the same PU has zero transition cost; switching from $P_j$ to $P_k$ adds D2H when leaving an accelerator and H2D when entering an accelerator. Edges from $v_{N,j}$ to $t$ contain the final D2H cost required by $P_j$. Each $s$--$t$ path therefore defines an operator-to-PU mapping whose total path cost predicts its end-to-end objective value. All reported results use a conservative instantiation of this graph that charges every operator its full H2D and D2H cost inside the node regardless of neighbor placement, upper-bounding transition costs (Section~\ref{subsec:ablation}).\looseness=-1

\noindent\textbf{Parallel execution graph.}\label{subsubsec:parallel_graph}
For general operator DAGs, \name uses topological traversal to partition execution into \textit{phases} bounded by fork and join points, and depth-first search to identify independent \textit{branches} within each phase. Operators within a branch execute sequentially, whereas data-independent branches may execute concurrently on separate PUs. Phase boundaries act as synchronization barriers, so a phase completes when its slowest contention-adjusted branch finishes; total latency is the sum of the phase makespans.\looseness=-1

\noindent\textbf{Memory contention modeling.}
For branches running on different PUs, \name multiplies each operator's latency by the measured slowdown factor
$\mathrm{SF}(P_{\text{run}},P_{\text{interfere}})$.
When multiple interfering PUs are active, it applies the maximum corresponding pairwise slowdown. On our platform, the NPU is most sensitive, slowing by $1.17\times$ with CPU interference and $1.09\times$ with GPU interference, while CPU--GPU interference is negligible. Two-request contention is instead captured by the directly measured co-execution costs described below. The pairwise model does not capture shape-dependent interference, PU-internal QoS effects, or non-pairwise three-PU contention.\looseness=-1

\noindent\textbf{Two-request concurrent inference.}
For two instances of the same model, \name constructs an aligned state space with states $(\ell,d_0,d_1)$, where $\ell$ is the completed operator step and $d_0,d_1$ are the current PU assignments. Both requests advance in lockstep, and each transition selects their next PU pair.

For different models, \name constructs joint states $(i,j)$, where $i$ and $j$ are the numbers of completed operators of each request. While both requests are unfinished, each transition advances both by one operator and selects the PU pair with the lowest co-execution cost. After one request finishes, the remaining request advances alone as a solo tail priced at profiled solo costs. In both modes, same-PU transitions use directly measured, barrier-synchronized co-execution costs, the pair makespan for mixed-model pairs and the symmetric average for same-model pairs, and cross-PU transitions use measured cross-PU co-execution latencies where profiled and the maximum of the two solo latencies otherwise. The concurrent search charges no separate transfer costs, since every profiled operator cost already includes its full H2D and D2H transfer.\looseness=-1

\noindent\textbf{Sequential search.}\label{subsubsec:sequential_search}
For $N$ operators and $K$ PUs, Dijkstra's algorithm~\cite{dijkstra} exactly minimizes the $s$--$t$ path cost in $O(NK^2\log(NK))$ time. Because the execution graph is acyclic, topological dynamic programming can instead solve it in $O(NK^2)$ time.\looseness=-1

\noindent\textbf{Intra-model parallel search.}\label{subsubsec:parallel_search}
Within each phase, \name applies the sequential search independently to every branch, allowing operator-level PU selection within each branch. The selected operator costs are adjusted by the maximum slowdown induced by the PUs active on the other branches; the phase latency is the maximum adjusted branch cost, and total latency is the sum of phase latencies. While sequential-chain search is globally optimal under the profiled cost model, the parallel method is a phase-wise heuristic, since branches are selected independently rather than searched jointly.\looseness=-1

\noindent\textbf{Two-request concurrent search.}\label{subsubsec:concurrent_search}
For two concurrent requests, Dijkstra searches the aligned or joint state space defined in Section~\ref{subsubsec:parallel_graph}. The resulting schedule is optimal within the selected state space under the measured co-execution cost model. The state-space restriction keeps the search tractable for the typical case of $K=2$--$3$ PUs.\looseness=-1

\noindent\textbf{Energy-optimal scheduling.}\label{subsubsec:energy_search}
Energy mode replaces each operator's total cost with measured package energy $e=t\,\bar{p}_j$, where $t$ is the operator's total profiled cost on PU $P_j$ and $\bar{p}_j$ is the model's sustained package power on that PU (Section~\ref{subsec:profiling}). The shortest path then minimizes total predicted package energy. In parallel phases, the contention-adjusted latency raises the corresponding energy. For two concurrent requests, the additive model prices each request with its solo power, an approximation because package power under co-execution is not measured directly.\looseness=-1

\subsection{Framework Overhead}
\label{subsec:overhead}

All profiling, graph construction, and search occur offline. A single-model graph contains $O(NK)$ nodes and $O(NK^2)$ edges. Even the largest evaluated model, a ${\sim}4{,}600$-operator VLA pipeline, yields fewer than $10{,}000$ feasible nodes. An aligned two-request graph has $O(NK^2)$ states and up to $O(NK^4)$ transitions, while a joint graph has $O(N_1N_2)$ states whose transitions each minimize over the $K^2$ PU pairs.

Search is lightweight in practice. Our unoptimized Python implementation solves the 448-operator Hyena single-model graph in 0.03\,s and the largest joint two-request graph, {$\pi_{0.5}$} paired with Hyena at ${\sim}2$M states, in 12\,s, so profiling dominates the one-time optimization cost. The resulting static schedule adds no optimization overhead to the inference critical path, although its execution incurs PU-boundary invocation costs quantified in Section~\ref{subsec:deployment}.\looseness=-1

\subsection{System Model}
\label{subsec:system_model}

\name assumes a static fused-operator DAG and fixed tensor shapes for each deployment configuration, with dynamic shapes requiring separate profiles or runtime selection among pre-profiled schedules. \name targets unified-memory edge SoCs, where the CPU, GPU, and NPU share off-chip DRAM~\cite{apple_m5_2025,intelcoreultra}. Consequently, profiled H2D and D2H costs capture cache-coherency, IOMMU, and DMA-setup overheads rather than physical tensor copies. These costs are 0.02--0.06\,ms/MB for tensors up to ${\sim}17$\,MB and reach ${\sim}0.18$\,ms/MB only at the largest measured 33\,MB size, consistent with spilling past the on-die caches (Fig.~\ref{fig:breakdown}(b)), so typical inter-operator tensors cross PU boundaries in tens of microseconds. Although the additive transition-cost model is approximate, Section~\ref{subsec:deployment} validates its end-to-end accuracy and shows that subgraph hand-off, rather than data movement, dominates boundary overhead.\looseness=-1

Inference frameworks such as OpenVINO compile kernels during initialization~\cite{openvino}, outside the inference critical path. Because \name relies on measured costs for an arbitrary PU set, porting it to another SoC requires only re-running the offline profiler, leaving the formulation and search unchanged.\looseness=-1

\section{Evaluation}
\label{sec:evaluation}

\subsection{Experimental Setup}
\label{subsec:setup}

\begin{figure}[t]
    \centering
    \includegraphics[width=\columnwidth]{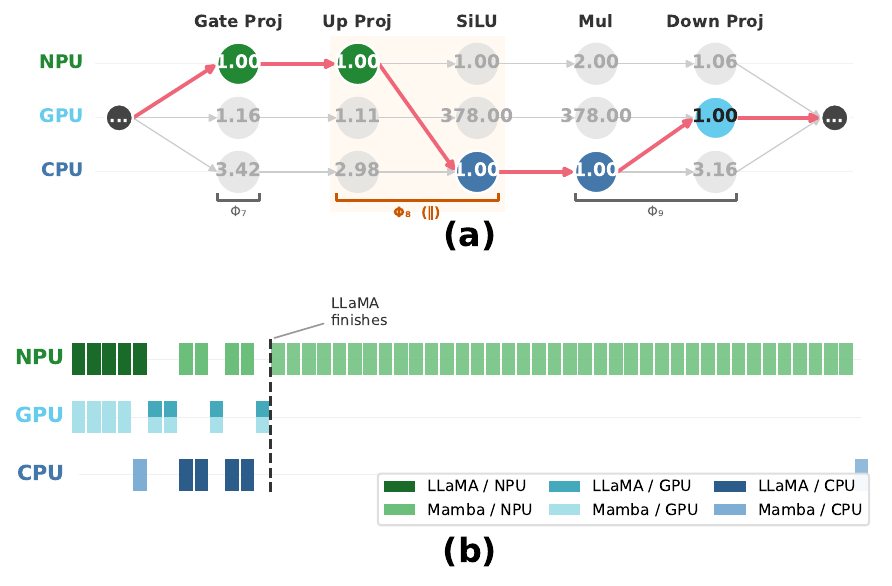}
    \vspace{-18pt}
    \caption{(a)~Sequential execution graph for five representative LLaMA-7B (1L) FP16 operators, annotated with execution phases (Section~\ref{subsubsec:parallel_graph}). Nodes show per-operator latency normalized to the best single-PU value, and the highlighted path is the selected mapping. $\Phi_7$--$\Phi_9$ are the 7th--9th phases of the full 13-operator graph, and $\Phi_8$ contains data-independent operators.
    (b)~Concurrent execution schedule for LLaMA-7B (1L) INT8 (13~fused ops) and Mamba-370M FP16 (52~fused ops).
    Each column is one step; row position indicates the assigned PU.
    The dashed line marks where LLaMA completes.}
    \label{fig:parallel_execution_graph}
    \vspace{-6pt}
\end{figure}

\noindent\textbf{Platform and workloads.}
We evaluate \name on an Intel Core Ultra (Lunar Lake, Series~2) SoC whose CPU, integrated GPU, and NPU share 16\,GB of memory, using OpenVINO~2025.3~\cite{intelcoreultra,openvino}. Table~\ref{tab:models} summarizes ten models spanning CNNs, Transformers, SSMs, VLAs, and emerging architectures, for 19 configurations. Nine models are evaluated at FP16 and NNCF-quantized INT8~\cite{nncf}; {$\pi_{0.5}$} uses one mixed-precision configuration.
Fused-operator counts differ across precisions because quantization changes the compiler's fusion boundaries.\looseness=-1

\begin{table}[t]
    \centering
    \caption{Evaluated models and fused-operator counts
    (FP16\,/\,INT8; {$\pi_{0.5}$}: one mixed-precision configuration).}
    \label{tab:models}
    \resizebox{\columnwidth}{!}{%
    \begin{tabular}{llcc}
    \hline
    \textbf{Model} & \textbf{Family} & \textbf{Input shape} & \textbf{Fused ops} \\
    \hline
    ResNet-50~\cite{resnet}      & CNN               & $1{\times}3{\times}224{\times}224$ & 73\,/\,68  \\
    ViT-B/16~\cite{vit}          & Vision Transformer& $1{\times}3{\times}224{\times}224$ & 175\,/\,151\\
    LLaMA-7B (1L)~\cite{llama}   & Transformer (one decoder layer) & $1{\times}128$      & 13\,/\,13  \\
    BitNet~\cite{bitnet}         & Ternary Transformer& $1{\times}128$                    & 36\,/\,35  \\
    Mamba-370M~\cite{mamba}      & Selective SSM     & $1{\times}128$                     & 52\,/\,75  \\
    Hyena~\cite{hyena}           & Long convolution  & $1{\times}1{\times}1024{\times}512$& 448\,/\,88 \\
    KAN~\cite{kan}               & Kolmogorov--Arnold& $1{\times}784$                     & 27\,/\,28  \\
    SNN-VGG9~\cite{snn-vgg}      & Spiking NN (25 steps)& $1{\times}1{\times}32{\times}32$& 93\,/\,97  \\
    LAVISH~\cite{lavish}         & Audio-visual      & dual: $224^2$\,+\,$128^2$         & 14\,/\,35  \\
    {$\pi_{0.5}$}~\cite{phi05}   & VLA (4 stages)    & multi-stage                        & 4595       \\
    \hline
    \end{tabular}}
    \vspace{-6pt}
\end{table}

\noindent\textbf{Measurement and accounting.}
Following Section~\ref{subsec:profiling}, each supported fused operator is measured with 200 timed iterations after 20 warm-up iterations. Transfer latencies are measured with Intel VTune~\cite{vtune}, dispatch latencies with system event tracing (ETW), and package power with HWiNFO64~\cite{hwinfo} during sustained inference, from which energy is derived as power~$\times$~latency. The three PUs share one package power domain, so package energy is exactly the quantity the energy objective minimizes. Unless otherwise noted, reported latency and energy results are composed from these profiles under identical accounting for \name and all baselines; Section~\ref{subsec:deployment} validates the resulting predictions end to end.\looseness=-1

\noindent\textbf{Baselines.}
For single-model execution, \textit{CPU-only}, \textit{GPU-only}, and \textit{NPU-only} run the complete model on one supported PU. Latency speedup is measured against the fastest single-PU execution for each configuration, whereas energy reduction is measured against the lowest-energy single-PU execution. For two-request experiments, \textit{homogeneous serial} executes both requests back to back on the single PU that minimizes their combined latency, the best schedule available without a second PU. The \textit{naive multi-PU} baseline instead runs both models concurrently as indivisible units and exhaustively selects the best whole-model PU pair; it therefore captures the benefit of additional PUs without operator-level orchestration. We further compare against OpenVINO \textit{HETERO}~\cite{openvino}, which assigns each supported subgraph to the first device in a priority list and is granted the best-performing ordering per configuration; a three-PU HaX-CoNN reimplementation~\cite{haxconn} with measured model-level contention factors and at most one intra-model PU switch; and an AxoNN reimplementation~\cite{axonn} that maps contiguous layer blocks across the two fastest PUs under an energy objective. All heterogeneous baselines use the same profiled cost tables.\looseness=-1

\subsection{Sequential Orchestration}
\label{subsec:sequential_results}

Fig.~\ref{fig:parallel_execution_graph}(a) illustrates \name's operator-level mapping for five representative LLaMA-7B (1L) FP16 operators. The optimal path assigns the projection operators across the NPU and GPU while placing lightweight non-linear operators on the CPU, demonstrating that PU affinity can vary even within a single layer.\looseness=-1

\begin{figure}[t]
    \centering
    \includegraphics[width=\columnwidth]{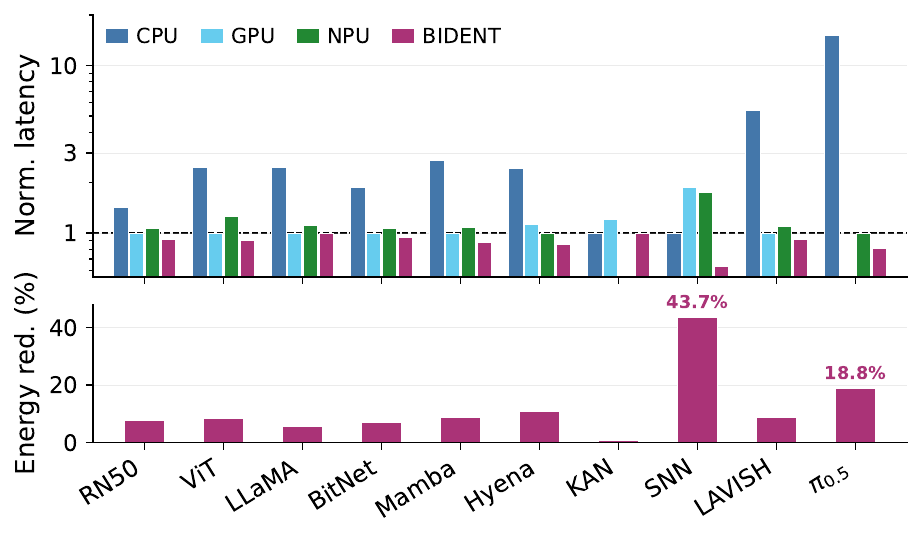}
    \vspace{-20pt}
    \caption{Sequential \name orchestration for the FP16 configurations ({$\pi_{0.5}$}: mixed; INT8 summarized in the text). Top: latency normalized to each model's best single PU (dashed line). Missing bars denote unsupported PUs. Bottom: energy reduction under energy-optimal scheduling vs.\ the best single-PU energy, whose PU may differ from the latency-best one.\looseness=-1}
    \label{fig:sequential_results}
    \vspace{-6pt}
\end{figure}

\begin{figure*}[t]
    \centering
    \includegraphics[width=\textwidth]{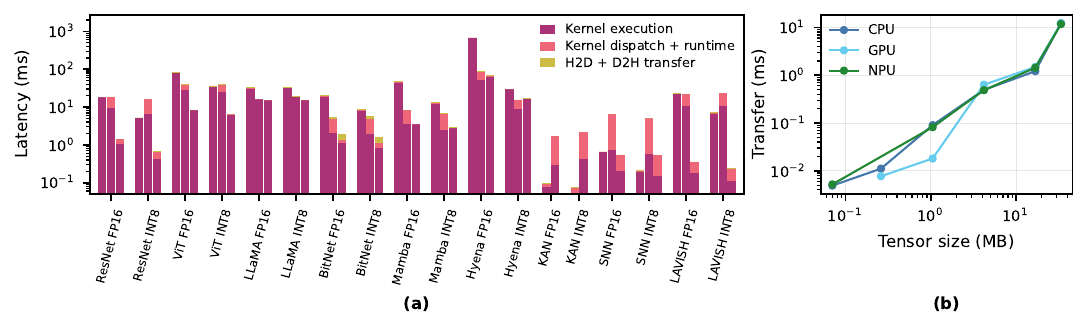}
    \vspace{-20pt}
    \caption{Absolute cost breakdown. (a)~Kernel-execution, kernel-dispatch, and host-transfer latency per model and PU for the 18 single-stage configurations. (b)~Host-boundary (H2D\,+\,D2H) transfer time versus tensor size, near-identical across PUs.}
    \label{fig:breakdown}
    \vspace{-6pt}
\end{figure*}

Fig.~\ref{fig:sequential_results} summarizes latency- and energy-optimized sequential orchestration for the FP16 configurations.\looseness=-1

\noindent\textbf{Latency.}
\name achieves 1.01--1.58$\times$ speedup across all configurations, with a geometric mean of 1.09$\times$. The largest gains occur in models with diverse operator affinities: SNN-VGG9 FP16 reaches 1.58$\times$ by distributing convolutions and lightweight spiking operations across the NPU and CPU; Mamba-370M FP16 and Hyena FP16 reach 1.14$\times$ and 1.16$\times$ by separating matrix operations from selective-scan or FFT-based operators; and {$\pi_{0.5}$} reaches 1.23$\times$ by mapping its heterogeneous pipeline stages across the CPU and NPU. Conventional or comparatively uniform workloads benefit less: ResNet-50, ViT-B/16, BitNet, and LAVISH improve by 1.04--1.10$\times$, while LLaMA and KAN remain within 1.02$\times$. The nine INT8 configurations, omitted from the figure, behave consistently, with 1.01--1.14$\times$ speedup and 0.2--13.7\% energy reduction. Sequential gains thus track a model's operator-level affinity diversity.\looseness=-1

\noindent\textbf{Energy.}
The energy-optimized schedule reduces energy for every configuration, averaging 9.2\% and reaching 43.7\% for SNN-VGG9 FP16, 18.8\% for {$\pi_{0.5}$}, and 10.9\% for Hyena FP16. The energy-optimized schedule's geometric-mean latency speedup is 1.03$\times$, compared with 1.09$\times$ for the latency-optimized schedule, reflecting the trade-off between the fastest and most energy-efficient PU. By contrast, latency optimization reduces energy by only 3.7\% on average and increases it for 7 of 19 configurations, two of them marginally, because it does not account for PU power. For example, the latency objective keeps KAN mostly on its fastest PU, the CPU, raising energy 12--18\% above the lowest-energy GPU execution, whereas the energy objective moves most KAN operators to the GPU. Energy must therefore be optimized from the profiled latency--power product.\looseness=-1

\noindent\textbf{Absolute cost breakdown.}
Fig.~\ref{fig:breakdown}(a) explains these results by separating kernel execution, dispatch, and host-transfer costs. CPU inference is execution-bound on 16 of the 18 single-stage configurations, whereas GPU dispatch and runtime overhead account for 33--89\% of latency on every model except the GEMM-dominated LLaMA layer. Host transfer is generally negligible, remaining below 1\% for the median configuration and becoming noticeable only for the shortest workloads. Accelerators thus excel on large dense operators but lose to the CPU on short or irregular kernels, motivating per-operator mapping.\looseness=-1

\begin{figure}[t]
    \centering
    \includegraphics[width=0.80\columnwidth]{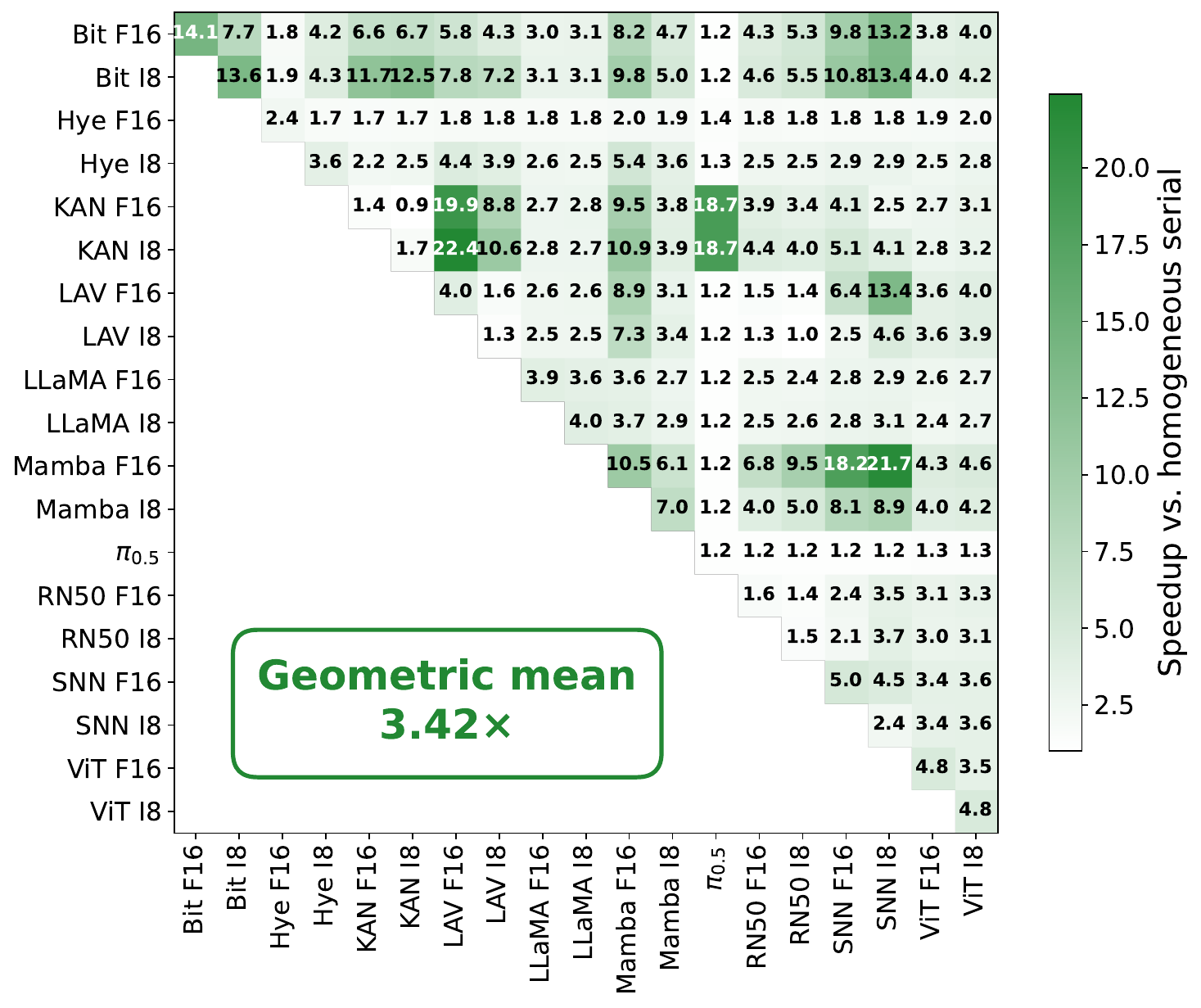}\\[3pt]
    \vspace{-5pt}
    \includegraphics[width=0.70\columnwidth]{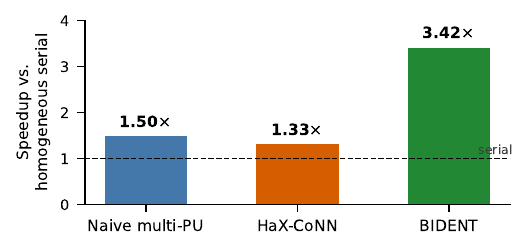}
    \vspace{-10pt}
    \caption{Concurrent orchestration over the 190 unique model pairs. Top: \name speedup per pair vs.\ homogeneous serial execution. Bottom: geometric-mean speedup of the naive multi-PU placement, HaX-CoNN~\cite{haxconn}, and \name over the same pairs and serial reference.}
    \label{fig:concurrent_heatmap}
    \vspace{-6pt}
\end{figure}

\subsection{Parallel and Concurrent Orchestration}
\label{subsec:parallel_results}

\noindent\textbf{Intra-model parallelism.}
On the FP16 configurations, parallel orchestration achieves 1.02--1.60$\times$ speedup over the best single PU, reducing latency by up to a further 29\% relative to the sequential \name schedule. Benefits track the number and balance of independent branches. ViT-B/16 and {$\pi_{0.5}$} gain 29\% and 16\% through 51 and 615 concurrent phases, Hyena, LAVISH, and Mamba-370M gain 8--12\%, while BitNet's linear chain offers no parallelism and the imbalanced branches of LLaMA and KAN yield at most 3\%. All reported makespans include measured cross-PU contention.\looseness=-1

\noindent\textbf{Parallel energy.}
Because the contention slowdown factor increases per-operator latency during concurrent phases, the energy-optimal parallel schedule saves less energy than the sequential one, 7.6\% versus 9.2\% averaged over the same 19 configurations.
The 1.6-point gap is the measured cost of memory bandwidth contention.
Models with many concurrent phases (ViT, Mamba, {$\pi_{0.5}$}) show the largest gap, and models with no concurrent phases (BitNet) show none. \looseness=-1

\noindent\textbf{Two-request concurrency.}
Fig.~\ref{fig:parallel_execution_graph}(b) shows LLaMA-7B and Mamba-370M co-scheduled across PUs, with Mamba continuing alone after LLaMA completes. Across all 190 model pairs, \name achieves a geometric-mean speedup of 3.42$\times$ over homogeneous serial execution (Fig.~\ref{fig:concurrent_heatmap}), ranging from 0.86$\times$ to 22.4$\times$. Speedups above 2$\times$ arise where the two models prefer different PUs, so any single shared PU is slow for at least one of them. This substantially exceeds the naive multi-PU placement (1.50$\times$) and HaX-CoNN (1.33$\times$), a 2.28$\times$ advantage over the naive placement gained through operator-level co-scheduling. Complementary-affinity pairs achieve the largest gains, including KAN$_{\mathrm{INT8}}$+\allowbreak LAVISH$_{\mathrm{FP16}}$ at 22.4$\times$, while same-model pairs such as BitNet$_{\mathrm{FP16}}$+\allowbreak BitNet$_{\mathrm{FP16}}$ still reach 14.1$\times$ by recovering idle-PU capacity. Bandwidth-intensive large-model pairs generally obtain smaller gains, and only 2 of 190 pairs regress because contention outweighs the available parallelism.\looseness=-1

\noindent\textbf{Concurrent energy.}
Energy-optimal concurrent scheduling reduces energy by 48.2\% on average relative to serial execution on each pair's lowest-energy PU, using the additive solo-power model of Section~\ref{subsubsec:energy_search}. Sixteen pairs see an energy increase where the contention penalty exceeds the energy benefit. \looseness=-1

\subsection{Comparison with Heterogeneous Baselines}
\label{subsec:baseline_comparison}

Table~\ref{tab:baseline_comparison} compares \name with the heterogeneous baselines defined in Section~\ref{subsec:setup}, using identical profiled cost tables. Single reports the geometric-mean speedup over the 19 configurations against the best single PU (18 for HETERO, whose harness covers the single-stage models), Concurrent the geometric-mean speedup over the 190 pairs against homogeneous serial execution, and Energy the mean single-model reduction against the best single-PU energy. \looseness=-1

\begin{table}[t]
    \centering
    \caption{Comparison with heterogeneous baselines under identical profiled costs.}
    \label{tab:baseline_comparison}
    \resizebox{\columnwidth}{!}{%
    \begin{tabular}{lccc}
    \hline
    \textbf{System} & \textbf{Single} & \textbf{Concurrent} & \textbf{Energy red.} \\
    \hline
    OpenVINO HETERO~\cite{openvino} & 1.00$\times$ & N/A & $-$2.6\% \\
    HaX-CoNN~\cite{haxconn} (reimpl.) & 1.02$\times$ & 1.33$\times$ & $+$2.4\% \\
    AxoNN~\cite{axonn} (block, 2 PUs) & 0.99$\times$ & N/A & $+$3.3\% \\
    \name & \textbf{1.09$\times$} & \textbf{3.42$\times$} & \textbf{$+$9.2\%} \\
    \hline
    \end{tabular}}
    \vspace{-6pt}
\end{table}

\name provides the strongest result for every supported objective. HETERO merely matches the best single PU even under its best per-model ordering, and increases energy because its priority-based policy selects devices by operator support rather than measured latency or power. HaX-CoNN's single-switch constraint captures limited operator-level affinity, reaching only 1.02$\times$ for single-model and 1.33$\times$ for concurrent execution. Its concurrent schedules are priced under its own static per-device slowdown factors of 1.39--1.65$\times$, which apply even to contention-free placements, so its predicted speedup falls below the naive placement, which is priced without contention. The AxoNN baseline reduces energy by 3.3\% but sacrifices latency because its two-PU contiguous-block mapping cannot express fine-grained assignments. These results show that \name's gains arise from combining operator-level mapping with unrestricted multi-PU co-scheduling.\looseness=-1

TensorRT~\cite{tensorrt} targets Jetson-class SoCs with DLA, ONNX Runtime's~\cite{onnxruntime} execution-provider fallback reduces to the same capability-based mapping as HETERO, and HeteroLLM~\cite{heterollm} and Agent.xpu~\cite{agent.xpu} are LLM-serving systems for other SoCs without public implementations, so none supports a direct comparison. OpenVINO AUTO is also excluded because it targets throughput rather than single-request latency, reaching only a measured 0.15$\times$ single-request speedup.\looseness=-1

\subsection{Component Ablation}
\label{subsec:ablation}

\begin{figure}[t]
    \centering
    \includegraphics[width=\columnwidth]{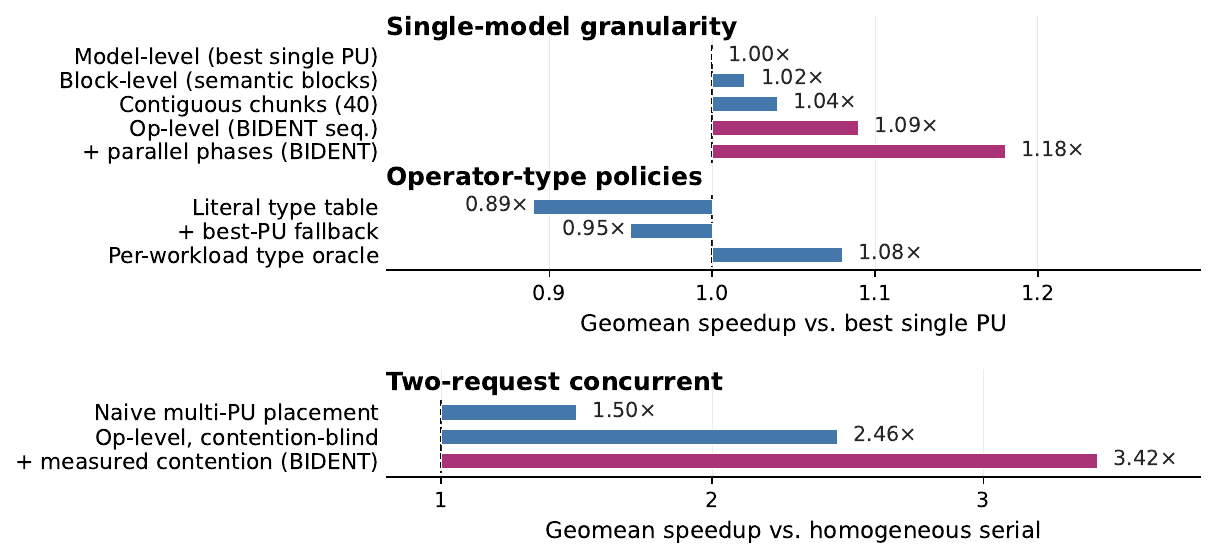}
    \vspace{-18pt}
    \caption{Component ablation. Bars extend from each block's baseline (dashed line).\looseness=-1}
    \label{fig:ablation}
    \vspace{-6pt}
\end{figure}

We isolate the effects of mapping granularity, operator-specific profiling, parallelism, and contention by optimizing each ablated variant under its own cost model and evaluating the resulting schedule under the full model (Fig.~\ref{fig:ablation}). The single-model and type-policy bars report geometric-mean speedup over the 19 configurations against the best single PU, and the concurrent bars over the 190 pairs against homogeneous serial execution.\looseness=-1

Granularity is the largest single-model contributor, and its gap widens with affinity diversity. SNN-VGG9 FP16 improves by 1.06$\times$ at block granularity but 1.58$\times$ at operator granularity. Under the conservative transfer accounting, which charges every operator its full H2D and D2H cost regardless of neighbor placement, a greedy per-operator mapper coincides with the sequential optimum, so granularity rather than solver sophistication is the operative axis. Static type-to-PU policies, built from the per-type winners in Fig.~\ref{fig:operator_type_comparison}, fare worse because they cannot capture size-dependent affinity. Even a per-workload type oracle stays below operator-level mapping, since 18\% of operators on average, and up to 38\%, prefer a different PU from their type-level choice, and the oracle would need the same per-workload profiling while expressing neither parallel phases nor co-scheduling. In the concurrent setting, measured contention modeling lifts contention-blind co-scheduling from 2.46$\times$ to 3.42$\times$ and changes the schedules of 187 of the 190 pairs, since the blind search prices same-PU co-execution at solo cost and over-shares PUs. Intra-model slowdown factors are instead applied after the search and change no assignments, but omitting them underestimates makespan by up to 9.5\%.\looseness=-1

\subsection{End-to-End Deployment Validation}
\label{subsec:deployment}

We deploy complete \name schedules through OpenVINO HETERO with manual per-operator device affinity, exercising real sub-model invocation, synchronization, and tensor hand-off. We re-profile the cost tables and re-derive the schedules on the deployment setup, keeping predictions and measurements on one basis, and measure package energy with RAPL. Validation covers 16 of the 19 configurations, excluding Hyena, whose manual-affinity deployments fall back to whole-model execution, and {$\pi_{0.5}$}, for which no deployment-side cost tables exist.\looseness=-1

Schedules that stay on one PU are predicted within 9\% (21\% for KAN, only 0.02\,ms in absolute terms), so the per-operator costs compose accurately. Multi-PU schedules pay a runtime cost that grows with the number of device boundaries, approximately 0.7\,ms per boundary over the 39 deployed schedules with a per-model median of 1.0\,ms, reflecting the executor's repeated sub-model invocation rather than mapping error. Affinity-diverse workloads retain their gains despite this overhead. SNN-VGG9 INT8 measures 1.57$\times$ faster than the best measured single PU and cuts energy from 6.47 to 1.83\,mJ, a 3.53$\times$ reduction, exceeding its original 1.14$\times$ predicted speedup because the re-profiled tables yield a different schedule. For models with small predicted gains, the boundary cost absorbs the benefit, motivating a resident executor.\looseness=-1

\section{Related Work}
\label{sec:related_work}

Prior heterogeneous inference systems address only parts of the orchestration problem. AxoNN~\cite{axonn} partitions CNN inference across CPU and DLA on NVIDIA Jetson with an energy-aware pipeline but is two-PU and CNN-specific, while HaX-CoNN~\cite{haxconn} models shared-memory contention across CPU, GPU, and DLA but targets inter-model scheduling. In contrast, \name jointly captures operator-level PU affinity, intra-model parallelism, and concurrent two-request execution across all three PUs of a modern SoC, including the NPU, within a profiling-driven graph formulation. Section~\ref{subsec:baseline_comparison} compares \name quantitatively against HaX-CoNN and AxoNN reimplementations on our platform.\looseness=-1

HeteroLLM~\cite{heterollm} partitions LLM prefill and decode across CPU, GPU, and NPU, while Agent.xpu~\cite{agent.xpu} extends heterogeneous scheduling to agentic LLM workloads. Although these systems demonstrate the value of heterogeneous execution, their scheduling policies are tailored to Transformer-specific phases and operators. \name instead applies the same profiling and search procedure across CNNs, Transformers, SSMs, KANs, Hyena, spiking networks, and multi-stage VLA pipelines without architecture-specific rules.\looseness=-1

FlexNPU~\cite{flexnpu} explores more flexible NPU datapaths and is complementary. Greater NPU flexibility reduces unsupported or inefficient operators but does not eliminate latency, energy, and parallelism differences among PUs, which \name addresses through measured operator-level scheduling rather than hardware modification.\looseness=-1
\section{Conclusion}
\label{sec:conclusion}

We present \name, a model-agnostic operator-level orchestration framework that uses profiled execution graphs and shortest-path search to coordinate sequential, intra-model parallel, and two-request concurrent inference across heterogeneous edge PUs. On an Intel Core Ultra SoC, \name achieves up to 1.58$\times$ sequential and 1.60$\times$ parallel single-model speedup, reduces sequential energy by 9.2\% on average, and delivers a 2.28$\times$ geometric-mean speedup over naive multi-PU placement across 190 concurrent model pairs. These results establish operator-level orchestration as a practical foundation for heterogeneous edge inference.\looseness=-1


\bibliographystyle{IEEEtran}
\bibliography{bib/trident}

\end{document}